%% file: spt3g_des_cmb_cluster_lensing.tex
\documentclass[a4paper,11pt]{article}

\pdfoutput=1 

\usepackage{jcappub} 
\usepackage{newtxtext,newtxmath}
\usepackage[T1]{fontenc} 
\usepackage{multirow}
\usepackage{wrapfig}
\usepackage{fp}
\usepackage{natbib}
\usepackage{color}
\usepackage{amsmath,amsfonts,bm}
\usepackage{hyperref}
\usepackage[T1]{fontenc}
\usepackage{graphicx}   
\usepackage{cleveref}
\usepackage[mathlines]{lineno}
\usepackage{booktabs}
\usepackage{amsmath}

\newcommand{\ukarcmin}{\mu{\rm K-arcmin}}
\newcommand{\uk}{\mu{\rm K}}
\newcommand{\sz}{Sunyaev–Zel{'}dovich}

\title{Mass calibration of DES Year-3 clusters via SPT-3G CMB cluster lensing}

\collaboration{SPT-3G \& DES Collaborations}

\def\Melbourne{1}
\def\ILNCSA{2}
\def\CTIO{3}
\def\Cardiff{4}
\def\LIneA{5}
\def\UMICH{6}
\def\FNAL{7}
\def\KICPChicago{8}
\def\AAUChicago{9}
\def\ILAst{10}
\def\IAP{11}
\def\ANLHEP{12}
\def\KIPAC{13}
\def\Stanford{14}
\def\SLAC{15}
\def\EFIChicago{16}
\def\PhysicsUChicago{17}
\def\Berkeley{18}
\def\KEK{19}
\def\McGill{20}
\def\CIFAR{21}
\def\PhysicsPrinceton{22}
\def\ColoradoAPS{23}
\def\ILPhys{24}
\def\UCLA{25}
\def\CaseWestern{26}
\def\UCDavis{27}
\def\CASA{28}
\def\ColoradoPhys{29}
\def\LBNL{30}
\def\Dunlap{31}
\def\UToronto{32}
\def\ANLMSD{33}
\def\Caltech{34}
\def\ThreeSpeedLogic{35}
\def\CfA{36}
\def\MSU{37}
\def\LMU{38}
\def\INNS{39}
\def\SORB{40}
\def\UCL{41}
\def\IAC{42}
\def\ULL{43}
\def\IFAE{44}
\def\IEEC{45}
\def\ICE{46}
\def\INAF{47}
\def\IFPU{48}
\def\HAMB{49}
\def\BRIS{50}
\def\CIEMAT{51}
\def\IIT{52}
\def\PENN{53}
\def\UGA{54}
\def\OSLO{55}
\def\UAM{56}
\def\SCIPP{57}
\def\CCAP{58}
\def\OSU{59}
\def\JPL{60}
\def\CWMI{61}
\def\LPSC{62}
\def\ICRE{63}
\def\MPE{64}
\def\OON{65}
\def\CMU{66}
\def\SUSS{67}
\def\UOS{68}
\def\CSMD{69}

\input{spt3g_des_cmb_cluster_lensing_authours}

\affiliation[\Melbourne]{School of Physics, University of Melbourne, Parkville, VIC 3010, Australia}
\affiliation[\ILNCSA]{Center for AstroPhysical Surveys, National Center for Supercomputing Applications, Urbana, IL, 61801, USA}
\affiliation[\CTIO]{Cerro Tololo Inter-American Observatory, NSF's NOIRLab, Casilla 603, La Serena, Chile}
\affiliation[\Cardiff]{School of Physics and Astronomy, Cardiff University, Cardiff CF24 3YB, United Kingdom}
\affiliation[\LIneA]{Laborat\'orio Interinstitucional de e-Astronomia - LIneA, Rua Gal. Jos\'e Cristino 77, Rio de Janeiro, RJ - 20921-400, Brazil}
\affiliation[\UMICH]{Department of Physics, University of Michigan, Ann Arbor, MI 48109, USA}
\affiliation[\FNAL]{Fermi National Accelerator Laboratory, P.O. Box 500, Batavia, IL, 60510, USA}
\affiliation[\KICPChicago]{Kavli Institute for Cosmological Physics, University of Chicago, 5640 South Ellis Avenue, Chicago, IL, 60637, USA}
\affiliation[\AAUChicago]{Department of Astronomy and Astrophysics, University of Chicago, 5640 South Ellis Avenue, Chicago, IL, 60637, USA}
\affiliation[\ILAst]{Department of Astronomy, University of Illinois Urbana-Champaign, 1002 West Green Street, Urbana, IL, 61801, USA}
\affiliation[\IAP]{Sorbonne Universit\'e, CNRS, UMR 7095, Institut d’Astrophysique de
Paris, 98 bis bd Arago, 75014 Paris, France}
\affiliation[\ANLHEP]{High-Energy Physics Division, Argonne National Laboratory, 9700 South Cass Avenue., Lemont, IL, 60439, USA}
\affiliation[\KIPAC]{Kavli Institute for Particle Astrophysics and Cosmology, Stanford University, 452 Lomita Mall, Stanford, CA, 94305, USA}
\affiliation[\Stanford]{Department of Physics, Stanford University, 382 Via Pueblo Mall, Stanford, CA, 94305, USA}
\affiliation[\SLAC]{SLAC National Accelerator Laboratory, 2575 Sand Hill Road, Menlo Park, CA, 94025, USA}
\affiliation[\EFIChicago]{Enrico Fermi Institute, University of Chicago, 5640 South Ellis Avenue, Chicago, IL, 60637, USA}
\affiliation[\PhysicsUChicago]{Department of Physics, University of Chicago, 5640 South Ellis Avenue, Chicago, IL, 60637, USA}
\affiliation[\Berkeley]{Department of Physics, University of California, Berkeley, CA, 94720, USA}
\affiliation[\KEK]{High Energy Accelerator Research Organization (KEK), Tsukuba, Ibaraki 305-0801, Japan}
\affiliation[\McGill]{Department of Physics and McGill Space Institute, McGill University, 3600 Rue University, Montreal, Quebec H3A 2T8, Canada}
\affiliation[\CIFAR]{Canadian Institute for Advanced Research, CIFAR Program in Gravity and the Extreme Universe, Toronto, ON, M5G 1Z8, Canada}
\affiliation[\PhysicsPrinceton]{Joseph Henry Laboratories of Physics, Jadwin Hall, Princeton University, Princeton, NJ 08544, USA}
\affiliation[\ColoradoAPS]{Department of Astrophysical and Planetary Sciences, University of Colorado, Boulder, CO, 80309, USA}
\affiliation[\ILPhys]{Department of Physics, University of Illinois Urbana-Champaign, 1110 West Green Street, Urbana, IL, 61801, USA}
\affiliation[\UCLA]{Department of Physics and Astronomy, University of California, Los Angeles, CA, 90095, USA}
\affiliation[\CaseWestern]{Department of Physics, Case Western Reserve University, Cleveland, OH, 44106, USA}
\affiliation[\UCDavis]{Department of Physics \& Astronomy, University of California, One Shields Avenue, Davis, CA 95616, USA}
\affiliation[\CASA]{CASA, Department of Astrophysical and Planetary Sciences, University of Colorado, Boulder, CO, 80309, USA }
\affiliation[\ColoradoPhys]{Department of Physics, University of Colorado, Boulder, CO, 80309, USA}
\affiliation[\LBNL]{Physics Division, Lawrence Berkeley National Laboratory, Berkeley, CA, 94720, USA}
\affiliation[\Dunlap]{Dunlap Institute for Astronomy \& Astrophysics, University of Toronto, 50 St. George Street, Toronto, ON, M5S 3H4, Canada}
\affiliation[\UToronto]{David A. Dunlap Department of Astronomy \& Astrophysics, University of Toronto, 50 St. George Street, Toronto, ON, M5S 3H4, Canada}
\affiliation[\ANLMSD]{Materials Sciences Division, Argonne National Laboratory, 9700 South Cass Avenue, Lemont, IL, 60439, USA}
\affiliation[\Caltech]{California Institute of Technology, 1200 East California Boulevard., Pasadena, CA, 91125, USA}
\affiliation[\ThreeSpeedLogic]{Three-Speed Logic, Inc., Victoria, B.C., V8S 3Z5, Canada}
\affiliation[\CfA]{Harvard-Smithsonian Center for Astrophysics, 60 Garden Street, Cambridge, MA, 02138, USA}
\affiliation[\MSU]{Department of Physics and Astronomy, Michigan State University, East Lansing, MI 48824, USA}
\affiliation[\LMU]{University Observatory, Faculty of Physics, Ludwig-Maximilians-Universit\"at, Scheinerstr. 1, 81679 Munich, Germany}
\affiliation[\INNS]{Universit\"{a}t Innsbruck, Institut f\"{u}r Astro- und Teilchenphysik, Technikerstr. 25/8, 6020 Innsbruck, Austria}
\affiliation[\SORB]{Sorbonne Universit\'es, UPMC Univ Paris 06, UMR 7095, Institut d'Astrophysique de Paris, F-75014, Paris, France}
\affiliation[\UCL]{Department of Physics \& Astronomy, University College London, Gower Street, London, WC1E 6BT, UK}
\affiliation[\IAC]{Instituto de Astrofisica de Canarias, E-38205 La Laguna, Tenerife, Spain}
\affiliation[\ULL]{Universidad de La Laguna, Dpto. Astrofísica, E-38206 La Laguna, Tenerife, Spain}
\affiliation[\IFAE]{Institut de F\'{\i}sica d'Altes Energies (IFAE), The Barcelona Institute of Science and Technology, Campus UAB, 08193 Bellaterra (Barcelona) Spain}
\affiliation[\IEEC]{Institut d'Estudis Espacials de Catalunya (IEEC), 08034 Barcelona, Spain}
\affiliation[\ICE]{Institute of Space Sciences (ICE, CSIC),  Campus UAB, Carrer de Can Magrans, s/n,  08193 Barcelona, Spain}
\affiliation[\ICE]{Astronomy Unit, Department of Physics, University of Trieste, via Tiepolo 11, I-34131 Trieste, Italy}
\affiliation[\INAF]{INAF-Osservatorio Astronomico di Trieste, via G. B. Tiepolo 11, I-34143 Trieste, Italy}
\affiliation[\IFPU]{Institute for Fundamental Physics of the Universe, Via Beirut 2, 34014 Trieste, Italy}
\affiliation[\HAMB]{Hamburger Sternwarte, Universit\"{a}t Hamburg, Gojenbergsweg 112, 21029 Hamburg, Germany}
\affiliation[\BRIS]{School of Mathematics and Physics, University of Queensland,  Brisbane, QLD 4072, Australia}
\affiliation[\CIEMAT]{Centro de Investigaciones Energ\'eticas, Medioambientales y Tecnol\'ogicas (CIEMAT), Madrid, Spain}
\affiliation[\IIT]{Department of Physics, IIT Hyderabad, Kandi, Telangana 502285, India}
\affiliation[\PENN]{Department of Physics and Astronomy, University of Pennsylvania, Philadelphia, PA 19104, USA}
\affiliation[\UGA]{Universit\'e Grenoble Alpes, CNRS, LPSC-IN2P3, 38000 Grenoble, France}
\affiliation[\OSLO]{Institute of Theoretical Astrophysics, University of Oslo. P.O. Box 1029 Blindern, NO-0315 Oslo, Norway}
\affiliation[\UAM]{Instituto de Fisica Teorica UAM/CSIC, Universidad Autonoma de Madrid, 28049 Madrid, Spain}
\affiliation[\SCIPP]{Santa Cruz Institute for Particle Physics, Santa Cruz, CA 95064, USA}
\affiliation[\CCAP]{Center for Cosmology and Astro-Particle Physics, The Ohio State University, Columbus, OH 43210, USA}
\affiliation[\OSU]{Department of Physics, The Ohio State University, Columbus, OH 43210, USA}
\affiliation[\JPL]{Jet Propulsion Laboratory, California Institute of Technology, 4800 Oak Grove Dr., Pasadena, CA 91109, USA}
\affiliation[\CWMI]{George P. and Cynthia Woods Mitchell Institute for Fundamental Physics and Astronomy, and Department of Physics and Astronomy, Texas A\&M University, College Station, TX 77843,  USA}
\affiliation[\LPSC]{LPSC Grenoble - 53, Avenue des Martyrs 38026 Grenoble, France}
\affiliation[\ICRE]{Instituci\'o Catalana de Recerca i Estudis Avan\c{c}ats, E-08010 Barcelona, Spain}
\affiliation[\MPE]{Max Planck Institute for Extraterrestrial Physics, Giessenbachstrasse, 85748 Garching, Germany}
\affiliation[\OON]{Observat\'orio Nacional, Rua Gal. Jos\'e Cristino 77, Rio de Janeiro, RJ - 20921-400, Brazil}
\affiliation[\CMU]{Department of Physics, Carnegie Mellon University, Pittsburgh, Pennsylvania 15312, USA}
\affiliation[\SUSS]{Department of Physics and Astronomy, Pevensey Building, University of Sussex, Brighton, BN1 9QH, UK}
\affiliation[\UOS]{School of Physics and Astronomy, University of Southampton,  Southampton, SO17 1BJ, UK}
\affiliation[\CSMD]{Computer Science and Mathematics Division, Oak Ridge National Laboratory, Oak Ridge, TN 37831}

\emailAdd{Behzad.ansarinejad@unimelb.edu.au}
\emailAdd{srinirag@illinois.edu}

\abstract{We measure the stacked lensing signal in the direction of galaxy clusters in the Dark Energy Survey Year 3 (DES Y3) redMaPPer sample, using cosmic microwave background (CMB) temperature data from SPT-3G, the third-generation CMB camera on the South Pole Telescope (SPT). Here, we estimate the lensing signal using temperature maps constructed from the initial 2 years of data from the SPT-3G `Main' survey, covering 1500 deg$^2$ of the Southern sky. We then use this lensing signal as a proxy for the mean cluster mass of the DES sample. The thermal Sunyaev–Zel'dovich (tSZ) signal, which can contaminate the lensing signal if not addressed, is isolated and removed from the data before obtaining the mass measurement. In this work, we employ three versions of the redMaPPer catalogue: a Flux-Limited sample containing 8865 clusters, a Volume-Limited sample with 5391 clusters, and a Volume\&Redshift-Limited sample with 4450 clusters. For the three samples, we detect the CMB lensing signal at a significance of $12.4\sigma$, $10.5\sigma$ and $10.2\sigma$ and find the mean cluster masses to be ${M}_{200{\rm{m}}}=1.66\pm0.13$ [stat.]$\pm 0.03$ [sys.], $1.97\pm0.18$ [stat.]$\pm 0.05$ [sys.], and $2.11\pm0.20$ [stat.]$\pm 0.05$ [sys.]$\times{10}^{14}\ {\rm{M}}_{\odot }$, respectively. This is a factor of $\sim2$ improvement relative to the precision of measurements with previous generations of SPT surveys and the most constraining cluster mass measurements using CMB cluster lensing to date. Overall, we find no significant tensions between our results and masses given by redMaPPer mass--richness scaling relations of previous works, which were calibrated using CMB cluster lensing, optical weak lensing, and velocity dispersion measurements from various combinations of DES, SDSS and Planck data. We then divide our sample into 3 redshift and 3 richness bins, finding no significant discrepancies with optical weak-lensing calibrated masses in these bins. We forecast a $5.7\%$ constraint on the mean cluster mass of the DES Y3 sample with the complete SPT-3G surveys when using both temperature and polarization data and including an additional $\sim1400$ deg$^2$ of observations from the `Extended' SPT-3G survey.}


\begin{document}
\begin{flushright}
\begin{tabular}{r}
\footnotesize FERMILAB-PUB-24-0123 \\
\footnotesize DES-2023-0821
\end{tabular}
\end{flushright}
\maketitle
\flushbottom

\section{Introduction}

Galaxy clusters are the most massive gravitationally collapsed objects and are the culmination of structure growth processes across cosmic time. As a result, cluster number counts as a function of cluster mass and redshift provide a sensitive probe of cosmological parameters that influence the growth of structure and the geometry of the Universe \cite[see reviews by][]{Allen2011, Weinberg2013}. These parameters include the matter density parameter, $\Omega_\textup{m}$; the normalisation of the matter power spectrum on the scale of $8\ h^{-1}$Mpc, $\sigma_8$; the dark energy equation of state parameter, $w$; as well as the sum of neutrino masses, $\sum m_\nu$ \cite[see e.g.][]{Schuecker2003, Salvati2018, Planck2016XXIV, Bocquet2019}. These constraints are highly complementary to those derived from analyses of baryon acoustic oscillations (BAO; \cite{Ansarinejad2018, Alam2021}), cosmic microwave background (CMB; \cite{Planck2018, Aiola2020, Balkenhol2022, Carron2022, Pan2023, MacCrann2024, Madhavacheril2024, Qu2024}), as well as auto- and cross-correlation analyses of optical weak gravitational lensing and galaxy clustering (3x2pt; \cite{DES2021, Heymans2020, Miyatake2023}) as these measurements have different parameter degeneracies and independent sources of systematics.

However, cosmological analysis of galaxy cluster samples is currently limited by our ability to reconstruct the mass distribution of the cluster sample (a problem called \emph{mass calibration}; see e.g. Section VI of \cite{Abbott2018} and Section 4 of \cite{Planck2016XXIV}, for a discussion of the impact of systematics on recent cluster cosmology analyses). In the near future, surveys such as eROSITA \cite{Merloni2012}, LSST \cite{LSST2019}, Euclid \cite{Laureijs2011}, Simons Observatory (SO; \cite{So2019}), and CMB-S4 \cite{Abazajian2016} will increase the cluster sample size compared to existing surveys by an order of magnitude, significantly reducing limitations due to statistical uncertainties. In preparation for these datasets, it is, therefore, crucial to improve our understanding of sources of systematic uncertainty that could impact commonly used cluster detection and mass calibration methods. In the optical regime, weak gravitational lensing is the most common cluster mass measurement approach \cite[see][for a review]{Umetsu2020}. Weak lensing offers the advantage of probing the total cluster mass with weak dependence on complex baryonic physics, which could affect the mass-observable scaling relations of the thermal \sz{} (tSZ) decrement, X-ray luminosity, and cluster richness. Weak lensing is, however, impacted by various sources of systematics error including bias in photometric redshift estimates, galaxy shape modelling errors and contamination of the lensed galaxy sample with cluster member or foreground galaxies \cite{Applegate2014, Simet2016, Melchior2017}. Calibration of these effects has led to a systematic floor smaller than $2\%$ on the halo mass \cite{Grandis2021b, Bocquet2023} if source galaxies can be reliably selected in the background of the cluster sample.

CMB cluster lensing is a promising alternative technique for measuring the masses of galaxy clusters. In this phenomenon, CMB photons passing through galaxy clusters' gravitational potential wells are deflected and due to the small-scale CMB gradient, form arcminute scale dipoles with amplitudes of $\lesssim 10 \uk$ \cite{Seljak2000, Hu2007}. Measurements of these dipoles can, therefore, be used as a proxy for the cluster mass. Furthermore, CMB cluster lensing and optical weak lensing have mostly independent systematics (though projection effects will impact both observables; see Section 4.2.5 for details), enabling us to verify whether these systematics have been correctly characterised and accounted for. Additionally, since the source plane of CMB lensing is the surface of last scattering at $z\sim1100$, mass calibration can also be carried out for higher cluster redshifts, where optical lensing starts to suffer from unreliable background source selection. This makes CMB cluster lensing an essential tool for cluster mass measurements in upcoming datasets such as CMB-S4 and SO, which greatly increase the size of high redshift cluster samples by detecting thousands of clusters at $z>1$ \cite{So2019, Raghunathan_CMBS4, raghunathan22}. As shown by \cite{Levy2023}, we expect mass constraints of $3.9\%$ and $1.8\%$ for a sample of 25,000 and 100,000 clusters detected by SO and CMB-S4 surveys respectively.   

Over the past two decades, several different methods have been developed to measure the CMB cluster and galaxy lensing signal from CMB temperature and polarization maps \cite{Holder2004, Maturi2005, Lewis2006, Hu2007, Yoo2010, Melin2015, Horowitz2019, Raghunathan2019Pol, saha24}. In recent years, a number of studies have obtained the first significant detections of CMB cluster lensing using data from various CMB experiments. Using CMB temperature data from the SPT-SZ survey conducted with the South Pole Telescope (SPT) and a sample of 513 clusters \cite{Bleem2015} detected with the same data, \cite{Baxter2015} obtained a $3.1\sigma$ measurement of the CMB cluster lensing signal. Using data from the Atacama Cosmology Telescope (ACT) and various galaxy samples, \cite{Madhavacheril2015} and \cite{Madhavacheril2020} obtained a $3.2\sigma$ and $4.2\sigma$ detection of the signal, respectively. Similar studies using the Planck CMB data include \cite{Planck2016XXIV} and \cite{huchet24}, where the lensing signal was measured at $\sim5\sigma$ for SZ-detected galaxy cluster samples, and \cite{Raghunathan2018}, where a sample of 12.4 million galaxies selected from the WISE and SCOS surveys were used as tracers of dark matter halos, obtaining a $17\sigma$ measurement of the lensing signal. 

Past studies have also applied CMB cluster lensing to calibrate the scaling relation between cluster mass and richness for different cluster samples detected with the redMaPPer algorithm \cite{Rykoff2014}. \cite{Geach2017} obtained a $10\%$ constraint on the richness--mass scaling relation using the Planck CMB data and optically detected clusters in the Sloan Digital Sky Survey (SDSS) data presented by \cite{Rykoff2014}. Later, \cite{Baxter2018} used the SPT-SZ CMB temperature map to obtain a $17\%$ constraint on the amplitude of the mass--richness scaling relation of redMaPPer clusters detected in the Dark Energy Survey \cite[DES][]{DES2005} Year 1 data \cite{McClintock2019}, while \cite{Raghunathan2019T} obtained a $\sim20\%$ measurement of the same scaling relation for the DES Year 3 (Y3) redMaPPer cluster sample, using CMB temperature maps from the SPTpol survey \cite{Austermann2012}. \cite{Raghunathan2019Pol} also used the same datasets to obtain the first detection of the CMB cluster lensing signal using only the polarization data, obtaining a $\sim 28\%$ mass constraint for richness $\lambda>10$ clusters in the DES Y3 sample. 

One of the main challenges in measuring the CMB cluster lensing signal is overcoming contamination due to various astrophysical foregrounds which could bias the cluster mass measurements if they are not accounted for. These sources of contamination include the tSZ and kinetic SZ (kSZ) effects, as well as the cosmic infrared background (CIB) due to the presence of dusty galaxies in the clusters (see \cite{Raghunathan2017} for a comprehensive discussion of the impact of these systematics on CMB cluster lensing measurements). Various techniques have been developed to overcome these sources of contamination, including cleaning the large-scale CMB gradient using a Quadratic Estimator (QE) to overcome the tSZ bias, as proposed by \cite{Madhavacheril2018}. \cite{Raghunathan2019} later developed a modified QE that overcame tSZ, kSZ, and CIB contamination by inpainting the large-scale gradient in the CMB maps to remove the cluster emission. \cite{Raghunathan2019Pol} presented a new estimator to measure the lensing dipole of stacked images, based on rotating cluster-centred CMB map cutouts along the direction of locally measured background CMB gradients. This approach is much simpler and less computationally expensive than the other alternative techniques and we adopt this method for measuring CMB cluster lensing in this study. We refer the reader to Section 5.2 of \cite{Levy2023}, for a performance comparison of the technique adopted in this work to other CMB cluster lensing estimators.

The layout of this paper is as follows. We present a summary of the galaxy cluster sample and the CMB maps used in our analysis in Section~\ref{sec: Datasets}, followed by a description of our methods and pipeline verification with simulations in Section~\ref{sec: pipeline description}. We then present our results and compare them with various other DES cluster mass measurements from the literature in Section~\ref{sec: Results and discussion}. We conclude by presenting a summary of our findings and their implications in Section~\ref{sec: Conclusions}. Throughout, we assume a $\Lambda$CDM cosmology with $h=0.6774$, $\Omega_m=0.307$, and $\Omega_{\Lambda}=0.693$. In this work, we express cluster masses as $M_{\rm200m}$, defined as the mass enclosed within a sphere whose average density is 200 times that of the mean matter density of the Universe, $\Bar{\rho}_m$, at the cluster redshift.

\begin{figure*}
    \centering
    \includegraphics[width=\textwidth]{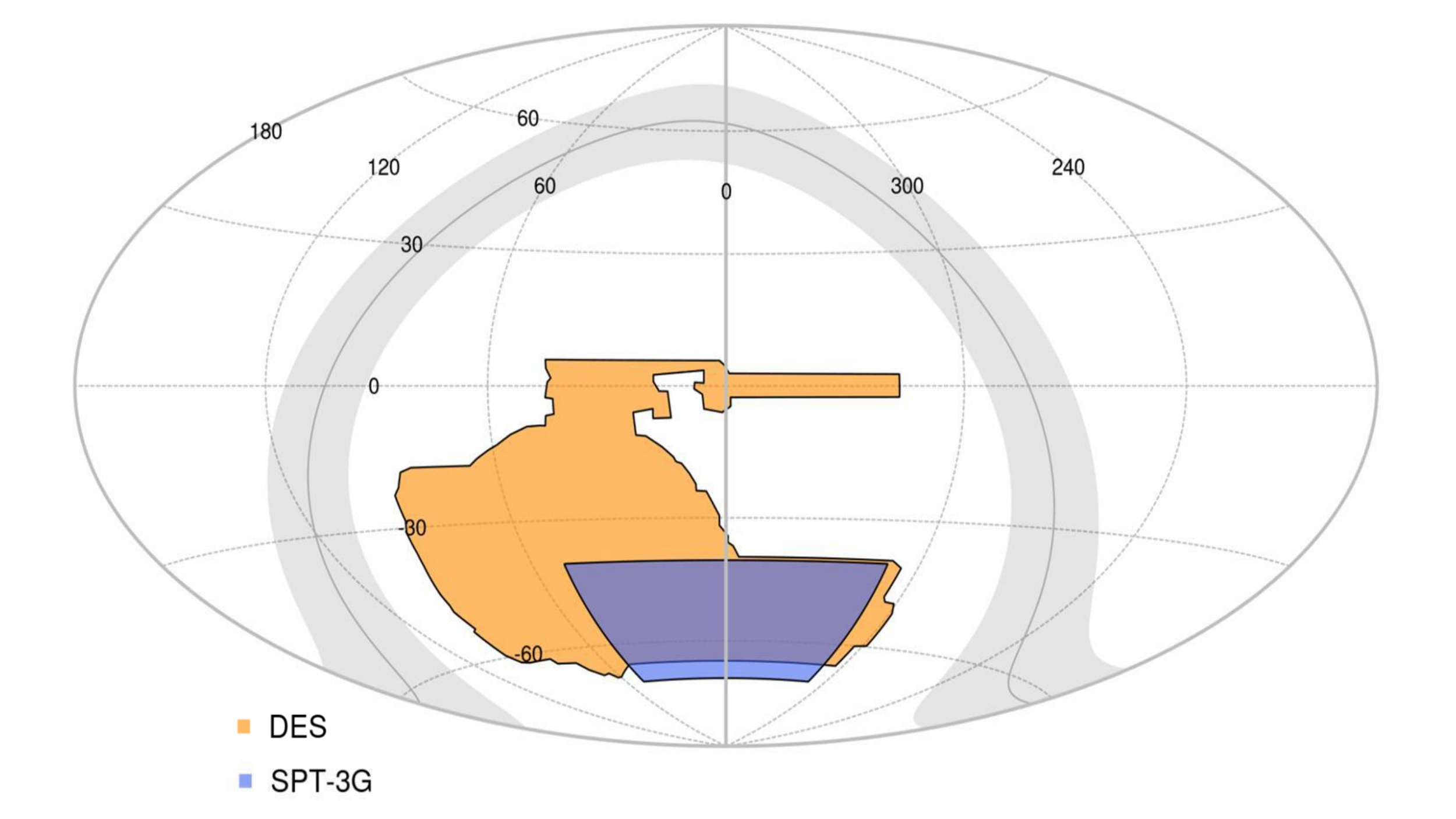}
    \caption{The $\sim1400$ deg$^2$ overlap between the $\sim1500$ deg$^2$ SPT-3G Main field (blue) and the $\sim5000$ deg$^2$ DES (orange) survey footprints. The grey band marks the galactic plane.}
    \label{fig:Survey_footprints}
\end{figure*}

\section{Datasets}
\label{sec: Datasets}

In this section, we provide a brief description of the datasets and sample selection used in our CMB cluster lensing analysis. 

\subsection{SPT-3G CMB data}
\label{sec:3G_data}

SPT is a 10-meter telescope located at the Amundsen-Scott South Pole station \cite{Carlstrom2011}, optimised for low-noise observations of the temperature and polarization of the CMB. SPT-3G \cite{Sobrin2022} is the third and latest receiver installed on the telescope, with the SPT-3G Main field covering a $\sim1500$ deg$^2$ footprint defined by $310^{\circ} < {\rm RA} < 50^{\circ}$ and $-70^{\circ} < {\rm DEC} < -42^{\circ}$. After masking point sources in the CMB data (see Section~\ref{sec:DES_Y3_RM_data}) and taking into account the masked area in the DES cluster sample, we are left with an overlap of $\sim1350$ deg$^2$ (see Figure~\ref{fig:Survey_footprints}) between the SPT-3G survey and the DES cluster catalogue. In this work, we use data from the initial two years of the SPT-3G survey observation (2019-2020). While the analysis of this paper is only performed using temperature data, we note that the inclusion of the initial two years of SPT-3G polarization data is expected to reduce the mass uncertainty by $\sim10\%$. We leave the polarization measurement to future works and provide forecasts for mass constraints using temperature and polarization data from the full survey depth, as well as data from the SPT-3G `Extended' survey (which provides an additional $\sim1400$ deg$^2$ of overlap with DES albeit at a lower sensitivity) in Section~\ref{sec: Forecasts}. In addition to the Main and Extended SPT-3G surveys, an additional $\sim6000$ deg$^2$ of the Southern sky will be observed for one year in the SPT-3G `Wide' survey. However, due to the relatively small additional overlap area with DES and the higher noise levels of these data, the Wide survey observations will not provide a significant improvement in the S/N of the CMB cluster lensing measurements, and we do not include these in our forecasts.

SPT-3G Main survey observations are conducted in the 95, 150, and 220 GHz bands with $1.6', 1.2'$, and $1.0'$ full width at half maximum beams and white noise levels of $\sim5$, 4, and 15 $\ukarcmin$ for the first two years of observations, in the three bands respectively. For each SPT-3G detector, the raw data are composed of digitised time-ordered data (TOD) that are converted to CMB temperature units (for details of the SPT-3G map making and data processing, see \cite{Dutcher2021}). During map making, we apply a $300<\ell_x<13000$ bandpass filter to the TOD. In this analysis, we use maps based on a minimum-variance combination of the 95, 150, and 220 GHz data, with Sanson-Flamsteed flat-sky projection \cite{Calabretta2002} and $0.5'$ pixels. 

\subsection{tSZ-nulled SPT-3G map}
\label{sec:null_tsz_map}

As described in Section~\ref{sec: Lensing estimator} our lensing estimator requires the estimation of the local CMB gradient at the location of each galaxy cluster. For this purpose, we use a tSZ-nulled CMB map constructed by performing an internal linear combination (ILC) of 95, 150, and 220 GHz data. As this map does not contain the tSZ signal, it allows for a more accurate estimation of the CMB gradient direction and amplitude, which in turn improves the S/N of our lensing dipole measurement. We note that not nulling the tSZ in the gradient estimation step increases the noise level when estimating gradient for our dataset.

\subsection{DES Y3 redMaPPer galaxy clusters}
\label{sec:DES_Y3_RM_data}

DES has a $\sim5000$ deg$^2$ footprint with imaging taken in the $g, r, i, z,$ and $Y$ bands via the Dark Energy Camera \cite{Flaugher2015} installed on the 4m Blanco telescope at the Cerro Tololo Observatory. The survey has completed the sixth and final year of observations, and here we use cluster samples detected using data from the first 3 years of the survey \cite{Sevilla-Noarbe2021}. We refer the reader to \cite{Rykoff2016} for a description of the application of the redMaPPer algorithm to the DES survey. In this work, we perform our analysis based on three redMaPPer cluster samples: a Flux-Limited sample, a Volume-Limited sample, and the Volume-Limited sample limited to the redshift range $0.2<z<0.65$. Henceforth, we shall refer to the latter as the Volume\&Redshift-Limited sample. While the Flux-Limited sample contains a significantly higher number of clusters at $z>0.65$, which would yield a lensing measurement with a higher S/N, the Volume-Limited sample is limited to varying redshifts which are determined based on the magnitude limit of the observations across the survey footprint. The Volume\&Redshift-Limited sample is then created to ensure sample uniformity across the survey footprint and match the selection applied to the sample used for the DES Y3 cluster cosmology analysis. As this sample has a well-understood selection function and is used for the DES cosmological analyses, we use the Volume\&Redshift-Limited as our baseline sample and the primary focus of the studies presented in subsequent sections. 

\begin{figure*}
        \centering
        \includegraphics[width=\textwidth]{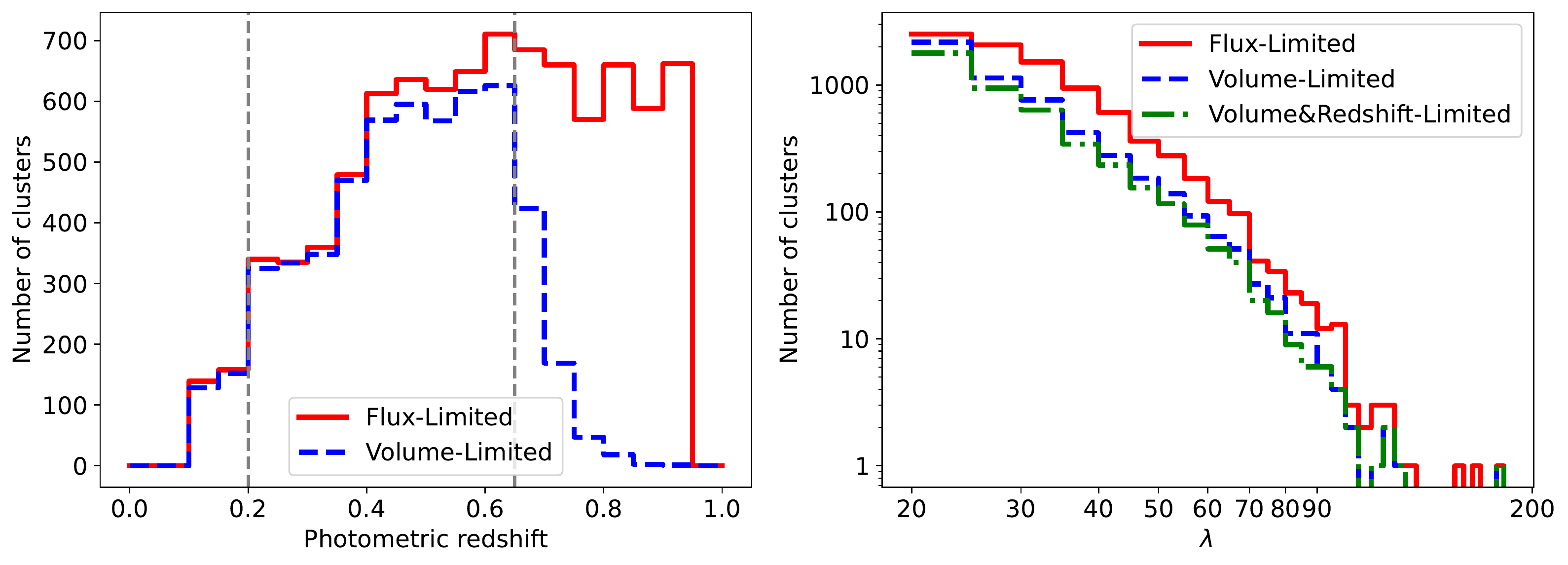}
    \caption{{\it Left panel:} The photometric redshift distribution of the DES Y3 flux and Volume-Limited redMaPPer cluster samples. The vertical dashed lines indicate the redshift cuts applied to the Volume-Limited sample. Here, we can see that most additional clusters in the Flux-Limited sample relative to the Volume-Limited sample are at $z>0.6$. {\it Right panel:} The richness distribution of the Flux-, Volume- and Volume\&Redshift-Limited samples. Although Volume-Limited samples contain fewer clusters, they follow a richness distribution similar to the Flux-Limited sample.}
    \label{fig:DES_lambda_nz}
\end{figure*}

In all cases, samples are limited to clusters with richness $\lambda\geq20$. Upon masking clusters within $1^{\circ}$ of the edges of the SPT-3G footprint and within 10$'$ of bright point sources ($\geq6$ mJy at 150GHz), we are left with 4450, 5391, and 8865 clusters in the Volume\&Redshift-Limited, Volume-Limited, and Flux-Limited samples, respectively. 
The photometric redshift and richness distributions of the three samples are shown in Figure~\ref{fig:DES_lambda_nz}. 
The median photometric redshifts of the Volume- and Flux-Limited samples are $\sim0.47$ and $\sim0.61$, with median uncertainties of $\sigma_z/(1+z)=0.006$ and $0.008$, respectively. 

To investigate the potential redshift or richness dependence of the cluster mass--richness scaling relation, we divide the Volume\&Redshift-Limited sample into 3 redshift and richness bins with an approximately equal number of clusters, as shown in Table~\ref{tab:z_lambda_subsamples}. 

\begin{table}
\centering
\caption{The number of clusters in the 3 redshift and richness subsamples of our Volume\&Redshift-Limited sample. The bins were chosen to have approximately equal numbers of clusters.}
\label{tab:z_lambda_subsamples}
\begin{tabular}{c|c}
\hline
bin definition & clusters per bin\\
\hline
\hline
$0.20<z<0.40$ & $1477$ \\

$0.40<z<0.53$ & $1480$ \\

$0.53<z<0.65$ & $1493$ \\
\hline
\hline
$20<\lambda<24$ & $1491$ \\

$24<\lambda<32$ & $1530$ \\

$32<\lambda$ & $1429$\\

\end{tabular}
\end{table}

\section{Pipeline description and validation}
\label{sec: pipeline description}

\subsection{Lensing estimator}
\label{sec: Lensing estimator}

We adopt the lensing estimator introduced by \cite{Raghunathan2019Pol}, which is briefly described here. Using the tSZ-nulled map described in Section~\ref{sec:null_tsz_map}, the algorithm first extracts $60'\times60'$ cutouts, centred on the location of $N_{\textup{clus}}$ clusters and $N_{\textup{rands}}$ random locations. The code then determines the median gradient direction $\theta_{\triangledown}=\tan^{-1}(\triangledown_y/\triangledown_x)$ from the central $6'\times6'$ region of each cutout. The noise penalty in the gradient estimation is reduced by applying a Wiener filter given by:

\begin{equation}
W_{\ell}=    
\begin{cases}
      C_{\ell}(C_{\ell}+N_{\ell})^{-1}, & \ell \leq2000\\
      0, & \text{otherwise}\\
\end{cases},
\label{eq:Wiener_filter}
\end{equation} to the $60'\times60'$ cutouts, where $C_{\ell}$ and $N_{\ell}$ are the data and noise power spectra, with the latter calculated using half-difference maps. 

Central $10'\times10'$ cutouts, ${\mathbf d}$, are then extracted from the SPT-3G map and rotated along the direction of the gradients, allowing for the stacking of the lensing dipoles which are oriented along the direction of the local CMB gradient.\footnote{Note that the Wiener filter is only applied to the larger cutouts of the tSZ-free map used for the gradient estimation step and the final $10'\times10'$ rotated cutouts are extracted from an unfiltered SPT-3G map.} At this stage, a weight is assigned to each cluster given by \mbox{$w=w_n w_g$}, where the $w_n$ component is based on the inverse noise variance $\sigma^2$ at the location of the cluster. The weight, $w_g$, is based on the median magnitude of the local gradient $\sqrt{\triangledown^2_{x}+\triangledown^2_{y}}$, which serves to maximize the $S/N$ of the measured dipole amplitude given its proportionality to the gradient amplitude.

\begin{linenomath*}
The cutouts are then mean subtracted\footnote{This mean subtraction will ensure the mean of the pixels in the stacked cutout is equal to zero.} and stacked to obtain the weighted stacks $\mathbf{s_c}$ and $\mathbf{s_r}$ at the location of clusters and random points, respectively. $\mathbf{s_c}$ is dominated by the mean large-scale CMB gradient (henceforth we refer to this as the background), which is estimated by $\mathbf{s_r}$ and corrected to obtain the final stacked dipole. To ensure that our background estimation is not biased due to sample variance, we set the value of $N_{\textup{rands}}=10\times N_{\textup{clus}}$. We have tested increasing $N_{\textup{rands}}$, and have found larger random samples for background subtraction do not change the results significantly. Here, we use the DES Y3 random catalogues which ensure that random points do not fall in the masked DES regions. The final stacked dipole is given by:
\end{linenomath*}

\begin{equation}
\mathbf{s} \equiv \mathbf{s_c}-\mathbf{s_r}=\frac{\sum^{N_{\textup{clus}}}_c w_c [\mathbf{d_c}-\langle \mathbf{d_c}\rangle]}{\sum^{N_{\textup{clus}}}_c w_c}-\frac{\sum^{N_{\textup{rand}}}_r w_r  [\mathbf{d_r}-\langle \mathbf{d_r}\rangle]}{\sum^{N_{\textup{rand}}}_r w_r}, 
\label{eq:stack_equation}
\end{equation}where $\mathbf{s}$ contains the dipole signal along with noise contributions from astrophysical and atmospheric foregrounds, instrumental noise, residual large-scale CMB gradient, the kSZ effect and in the case of temperature maps, the tSZ effect. Here $w_c$ and $w_r$ are the weights ($w=w_n w_g$) at the cluster and random locations respectively.

\begin{figure*}
    \centering
    \includegraphics[width=\textwidth]{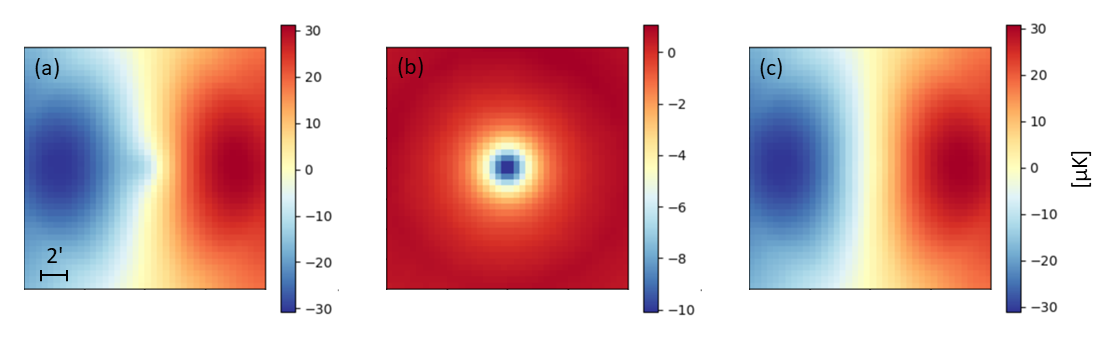}
    \caption{(a) The rotated and weighted cluster stack $\mathbf{s_c}$ from our minimum variance temperature map, including the tSZ contamination. (b) The mean of 25 randomly rotated cluster stacks, with the tSZ signal visible at the centre of the cutout. (c) Panel (a)-(b): the final data stack after removing tSZ contamination.}
    \label{fig:tsz_removal}
\end{figure*}

To remove the tSZ contribution from the stack (which could cause a bias toward lower masses if not taken into account), we follow the approach of \cite{Levy2023} and rotate the cluster cutouts in random orientations prior to stacking. To ensure an accurate estimate of the tSZ contamination, we repeat this procedure 25 times and take the mean of the 25 stacks as our estimate of the tSZ signal. We note that in addition to tSZ, this procedure also removes the CIB and all other cluster correlated foreground signals from our final stack. Fig.~\ref{fig:tsz_removal} illustrates the main steps involved in removing tSZ, with panel (a) showing our original stack (with tSZ contamination visible in the central pixels of the image), while panel (b) shows the mean of our 25 `random rotation' stacks with an estimate of tSZ contamination. Finally, panel (c) shows our tSZ-free cluster stack which is obtained by subtracting the tSZ signal shown in panel (b) from panel (a).

\subsection{Lensing dipole models}
\label{sec: Lensing dipole models}

To create the lensing dipole models $\mathbf{m}\equiv\mathbf{m}(M)$, we generate noiseless cluster-lensed simulations for a set of $N_{\textup{clus}}$ clusters with the redshift distribution of the DES sample and cluster masses varying in the range $M\in[0,4]\times10^{14}$\(\rm{M}_\odot\) with linear bins of $\Delta M=0.1\times10^{14}$\(\rm{M}_\odot\). For each mass bin, the $N_{\textup{clus}}$ cutouts are then stacked, following the steps in the previous section. Here, the mean background is simply given by $\mathbf{m_r}\equiv\mathbf{m_c}(M=0)$ and is subtracted from the stacks in all mass bins. Since the uncertainties on the gradient direction ($\delta \theta_{\triangledown}$) measurement will be lower in the case of noiseless simulations relative to the real data, the suboptimal stacking of the lensing dipole in the data compared to the models will cause a bias towards lower masses in the likelihoods, if not corrected for. For this reason, we add white noise and Gaussian foregrounds (mimicking those present in the data) to the models only when measuring $\delta \theta_{\triangledown}$. In Section~\ref{sec: Simulations}, we provide a more comprehensive description of the generated simulations, including the prescription for adding the impact of cluster miscentering and correlated structure to our modelled lensing profiles.

\subsection{Stacked cluster mass likelihood}
\label{sec: Mass likelihood}

Equipped with the stacked dipole signal, $\mathbf{s}$, and the models, $\mathbf{m}$, we calculate the likelihood using 
\begin{equation}
-2\ln{{\cal L}(M|\mathbf{s})}=\sum^{}_{\textup{pixels}}(\mathbf{s}-\mathbf{m})\hat{\mathbf{C}}^{-1}(\mathbf{s}-\mathbf{m})^T,
\label{eq:likelihood_equation}
\end{equation} where $\hat{\mathbf{C}}$ is the covariance matrix which is estimated from the data using the jackknife resampling technique by dividing the data into $N_{\textup{jk}}=0.9N_{\textup{clus}}$ subsamples:
\begin{equation}
\hat{\mathbf{C}}=\frac{N_{\textup{jk}}-1}{N_{\textup{jk}}}\sum^{N_{\textup{jk}}}_{i=1}[\mathbf{s}_i-\langle \mathbf{s}\rangle][\mathbf{s}_i-\langle \mathbf{s}\rangle]^T.
\label{eq:JK_covmat}
\end{equation}Here, $\mathbf{s}_i$ is the data stack in the $i$-th subsample and $\langle \mathbf{s}\rangle$ is the ensemble average of all subsamples. Estimating the covariance matrix from the data offers the advantage of capturing all sources of noise impacting the stacked lensing signal. When computing the likelihoods for our redshift and richness subsamples, which contain roughly one-third of the clusters from our Volume\&Redshift-Limited sample, we adopt the covariance matrix derived from the latter and multiply it by a factor of 3 to account for the increased shot noise in the subsamples. This approach sidesteps potential inaccuracies in estimating the covariance matrix that could result from using the smaller cluster samples from our redshift and richness subsamples.      

\subsection{Simulations \& pipeline validation}
\label{sec: Simulations}
\begin{linenomath*}
In order to test the pipeline and estimate the expected S/N of our measurements, we follow a similar approach to \cite{Raghunathan2019Pol} and create simulations of the lensed SPT-3G CMB temperature maps with properties similar to our minimum variance combination of the 95, 150, and 220 GHz maps from the real data. We generate a set of $N_{\textup{clus}}$ simulations by creating Gaussian realisations of the CMB in $60'\times60'$ flat-sky maps. For each cluster, we model the convergence profile as $\kappa_{\textup{tot}}(M,z)=\kappa_{\textup{1h}}(M,z)+\kappa_{\textup{2h}}(M,z)$. We model the one-halo term as a Navarro-Frenk-White \cite[NFW;][]{NFW1996} profile, with the concentration parameter given by \cite{Diemer2019}. We account for the impact of uncertainties due to cluster miscentering following \cite{Oguri2011}:
\begin{equation}
\Tilde{\kappa}_{\textup{1h}}(\ell)=\kappa_{\textup{1h}}(\ell)\biggl[(1-f_{\textup{mis}})+f_{\textup{mis}} \exp{\biggl(-\frac{1}{2}\sigma^2_s\ell^2\biggr)}\biggr]. 
\label{eq:one-halo}
\end{equation} Here, we use the DES miscentering fraction $f_{\textup{mis}}=0.22\pm0.11$ given by \cite{Rykoff2016} and $\sigma_s=\sigma_R/D_A(z)$, where $D_A(z)$ is the angular diameter distance at the cluster redshift. The magnitude of miscentering is modelled as a Rayleigh distribution with $\sigma_R=c_{\textup{mis}}R_{\lambda}$, where $R_{\lambda}=(\lambda/100)^{0.2}h^{-1}$Mpc is the DES redMaPPer cluster radius, and $\ln{c_{\textup{mis}}}=-1.13\pm0.22$ \cite[see][]{Melchior2017}. Following the approach of \cite{Raghunathan2019T}, we model the two-halo term, $\kappa_{\textup{2h}}$, which takes into account the contribution of correlated structures to the total lensing convergence, following Equation (13) of \cite{Oguri2011_2halo}.
\end{linenomath*}
We convolve the CMB maps with the SPT-3G 150 GHz beam function \cite[][]{Sobrin2022} and add the noise measured from our minimum variance combination of the 95, 150, and 220 GHz maps from the real data. We also include cluster tSZ and kSZ signals based on the Agora simulation set \cite{Omori2022} and include the foreground power due to CIB and radio galaxies based on measurements by \cite{George2015}. While studying the impact of alternative
simulations (e.g. Websky \cite{Stein2020} and Sehgal \cite{Sehgal2010} simulations)  is beyond
the scope of the current work, quantifying the simulation dependence would be an interesting topic of investigation for future works. To mimic the impact of filtering applied to the data during the map-making process, we follow previous works (including \cite{Raghunathan2019Pol}; \cite{Baxter2018}) and apply a 2D transfer function of the form $F_{\bar{\ell}}=e^{-(\ell_1/\ell_x)^6}e^{-(\ell_x/\ell_2)^6}$, with $\ell_1=300$ and $\ell_2=13000$ to the simulations. 

The central $10'\times10'$ region of each simulation is then extracted and processed through the pipeline as described in Sections~\ref{sec: Lensing estimator} and \ref{sec: Lensing dipole models} for the simulated data and models, respectively. For the purpose of pipeline verification, we generated 25 sets of mock CMB cluster lensed simulations each with $N_{\textup{clus}}$, one set of random CMB simulation with $N_{\textup{rand}}=10\times N_{\textup{clus}}$ and 25 sets of model simulations per mass bin, each with $N_{\textup{clus}}$ lensed clusters. In the case of the 25 mock simulation sets, we estimated a unique jackknife covariance matrix for each simulation set and set $N_{\textup{clus}}=5500$, roughly matching the number of clusters in the DES-Y3 Volume-Limited sample described in Section~\ref{sec:DES_Y3_RM_data}. For each simulated cluster, we assign the mass and redshift corresponding to a real cluster in the DES-Y3 sample. In the case of the models, where the cluster masses are fixed in each mass bin, we set the redshift distribution of the clusters to that of the DES-Y3 sample. To convert the cluster richness to cluster mass, we use the $M-\lambda$ scaling relation based on the weak lensing analysis of \cite{McClintock2019} given by $\langle M_{\rm 200m}|\lambda,z\rangle = 3.081\times10^{14}\ (\lambda/40)^{1.356}((1+z)/1.35)^{-0.3}$, resulting in a mean mass of $M_{\rm 200m}=2.21\times 10^{14}{\rm M}_\odot$ for the DES Volume\&Redshift-Limited cluster sample used in this analysis. We then evaluate the significance of our lensing measurement using $S/N=\sqrt{2[\ln{{\cal L}}(M_{\rm 200m}=M_{\textup{DES}})-\ln{{\cal L}}(M_{\rm 200m}=0)]}$, where $M_{\textup{DES}}$ is our measured stacked mass of the DES cluster sample.

\section{Results and Discussion}
\label{sec: Results and discussion}

\subsection{Pipeline verification results}
\label{sec: Pipeline_verification_results}

\begin{figure}
    \centering
    \includegraphics[width=0.75\columnwidth]{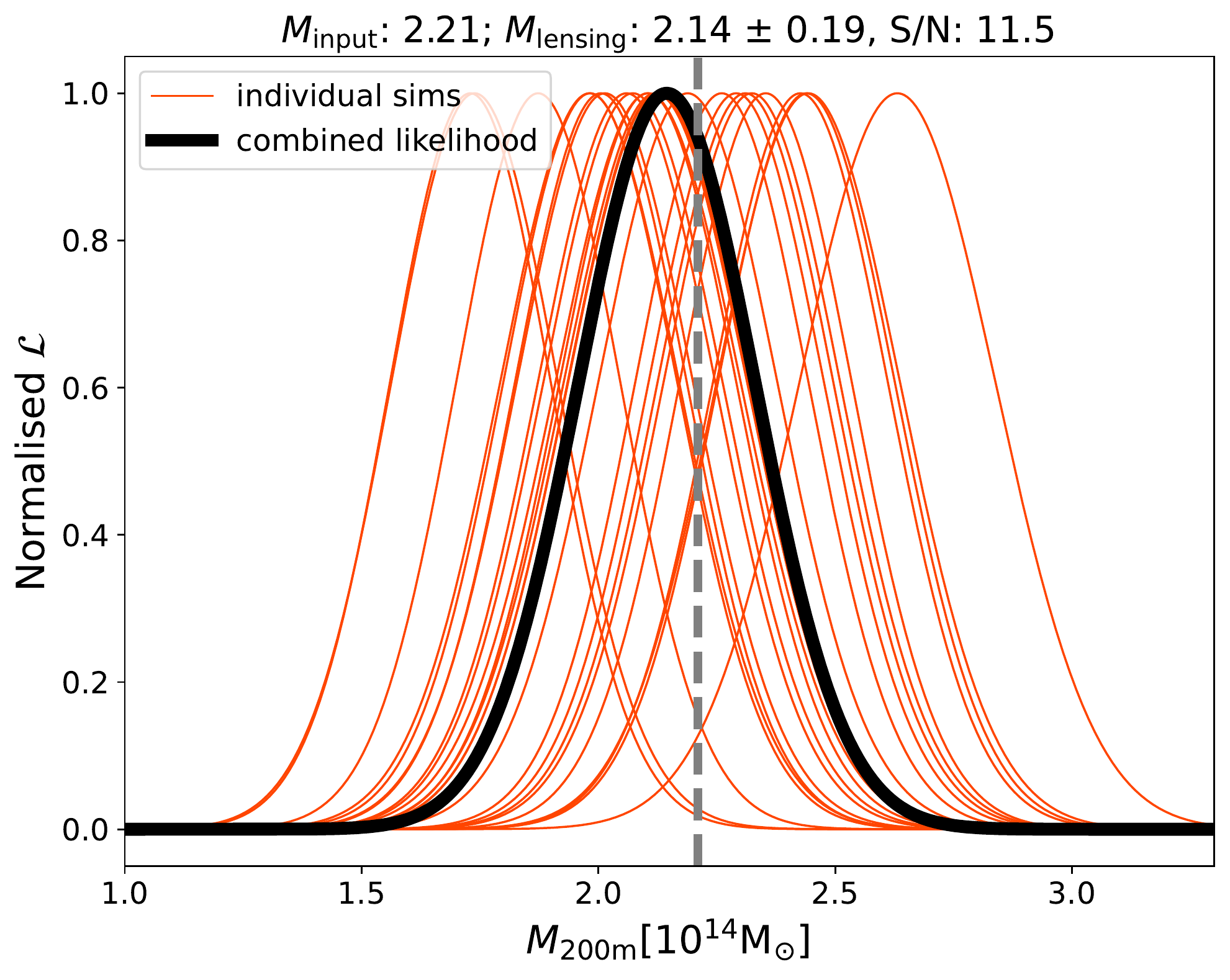}
    \caption{Likelihoods showing the results of the pipeline test for 25 sets of simulations (thin orange curves) described in Section~\ref{sec: Simulations} and their combined likelihood (thick black curves). For the combined likelihood, the width of the distribution is scaled up by a factor of 5 in order to demonstrate the expected S/N from one simulation run with the same number of clusters as our real data. The dashed vertical line is the mean cluster mass of the input sample used to produce the lensing simulations. We find a good agreement between the input mass and the recovered mass.}
    \label{fig:pipeline_test_likelihoods}
\end{figure}

Figure~\ref{fig:pipeline_test_likelihoods} shows the result of our pipeline verification test based on 25 sets of simulations (shown as thin orange curves). The thick black curve shows the combined likelihood with the standard deviation multiplied by a factor of $\sqrt{25}$, to estimate the S/N expected from a cluster dataset with the mass and redshift distribution of the DES Volume-Limited sample. The dashed vertical line represents the mean mass of the cluster sample ($2.21\times10^{14}$\(\textup{M}_\odot\)) and we find that the pipeline successfully recovers the expected mass of the input sample. 

The successful pipeline validation indicates that the intrinsic richness scatter of the DES sample does not have a significant impact on measuring the mean cluster mass of the sample. Furthermore, the pipeline test indicates the validity of the assumption that, by measuring the mean lensing signal, we can measure the mean cluster mass. 

\subsection{Mean cluster mass}
\label{sec_mass_richness_relation_norm}

\begin{linenomath*}
Our CMB cluster lensing measurement results in a mean (stacked) cluster mass of
\begin{align}
    {M}_{200{\rm{m}}} &= 1.66 \pm 0.13 \text{ [stat.]}\pm 0.03 \text{ [sys.]} \times 10^{14}\,\textup{M}_\odot \quad (8.0\%), \nonumber \\
    {M}_{200{\rm{m}}} &= 1.97 \pm 0.18 \text{ [stat.]}\pm 0.05 \text{ [sys.]} \times 10^{14}\,\textup{M}_\odot \quad (9.5\%), \nonumber \\
    {M}_{200{\rm{m}}} &= 2.11 \pm 0.20 \text{ [stat.]} \pm 0.05 \text{ [sys.]} \times 10^{14}\,{\rm{M}}_{\odot } \quad (9.8\%) \nonumber.
\end{align} for the DES-Y3 Flux-, Volume-, and Volume\&Redshift-Limited samples, respectively. These are the most constraining mass measurements obtained from CMB cluster lensing to date. Here, we find a good agreement ($<0.5\sigma$) between the masses of the Volume- and Volume\&Redshift-Limited samples, as expected given that they largely contain the same clusters (see left panel of Fig.~\ref{fig:DES_lambda_nz}). On the other hand, we find the stacked mass of the Flux-limited sample to be $\sim20\%$ lower (similar to the findings of previous studies \cite{Raghunathan2019T}), which is also in line with expectations given the higher number of low richness clusters in this sample compared to the volume-limited samples (see right panel of Fig.~\ref{fig:DES_lambda_nz}).
\end{linenomath*}

\begin{figure*}
    \centering
    \includegraphics[width=0.95\textwidth]{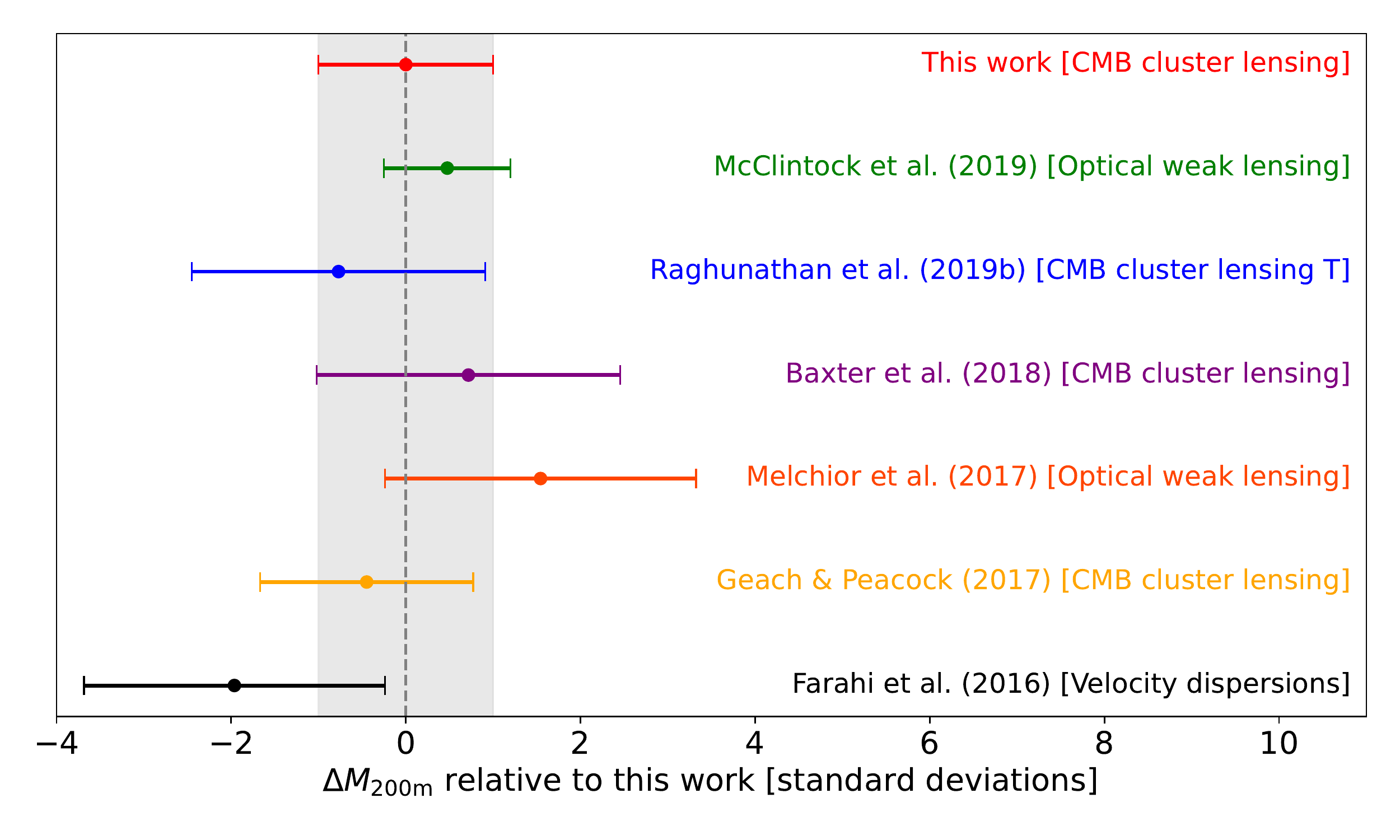}
    \caption{Comparison of the measured mean cluster mass of this work  (Volume\&Redshift-Limited sample) and the mean cluster mass of the same sample, based on the redMaPPer cluster mass--richness scaling relations of various other studies.}
    \label{fig:vol_lim_likelihoods}
\end{figure*}

Figure~\ref{fig:vol_lim_likelihoods} shows a comparison of the mean cluster mass for our Volume\&Redshift-Limited sample and the mean cluster mass of the same sample based on the mass--richness scaling relations of \cite{McClintock2019}, \cite{Melchior2017} (calibrated using optical weak lensing), \cite{Raghunathan2019}, \cite{Baxter2018} and \cite{Geach2017} (calibrated using CMB cluster lensing) and \cite{Farahi2016} (calibrated using spectroscopic velocity dispersion measurement of cluster galaxies). For ease of comparison, all measurements are normalised relative to this work with the shaded region marking the $1\sigma$ uncertainty on our measurement. Overall, we find a good agreement between our measurement of the mean cluster mass and the results of other studies.

\subsection{Systematics}
\label{sec: Systematics_tests}

In this section, we explore contributions from beam uncertainties, transfer function modelling, residual foregrounds, and cluster miscentering to the systematic error budget. We focus on these sources of systematics, as they have been shown to dominate systematic uncertainty in previous works (see Section 4.2 of \cite{Raghunathan2019T}). Here, we ignore systematic contributions from underlying cosmology and the choice of halo profile, as these have been found to be small in previous analyses (see \cite{Raghunathan2019T}, \cite{Baxter2018}) and negligible given the current magnitude of statistical uncertainty. However, quantification of these systematics will become important in future experiments such as CMB-S4, where the statistical uncertainty is expected to be reduced to $1\%$ given the much larger sample size of $\sim100,000$ clusters.  

\subsubsection{Cluster tSZ Signal and Residual Foregrounds}
\label{sec: tsz_fg_sys}

The tSZ signal is an important source of systematics, which could result in a bias towards lower masses if not correctly accounted for. In order to verify the success of the random rotation method (described in Section~\ref{sec: Lensing estimator}) in removing the tSZ signal from our stacked lensing dipole, we repeat the measurement using the tSZ-nulled minimum-variance map, described in Section~\ref{sec:null_tsz_map}, for our Volume\&Redshift-Limited sample.

\begin{figure}
    \centering
    \includegraphics[width=0.8\columnwidth]{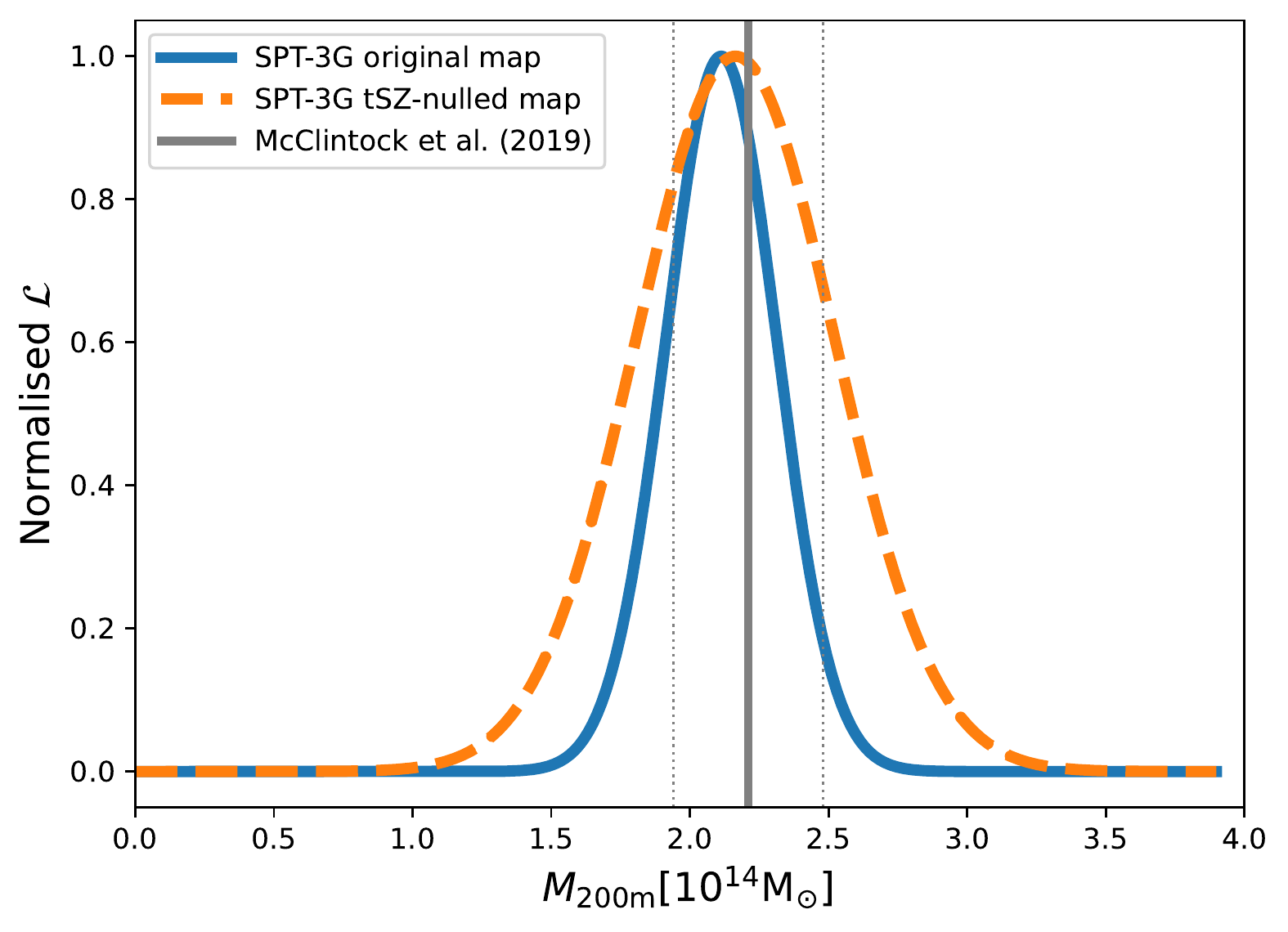}
    \caption{Comparison of mass likelihoods obtained using the SPT-3G tSZ-nulled temperature map (dashed orange curve) and our original maps (solid blue curve) in which the tSZ signal is estimated using random rotations and removed from the lensing stack. For comparison, we also include mean sample mass based on the optical weak lensing calibrated, mass--richness scaling relation of \cite{McClintock2019} (solid vertical line with the $1\sigma$ error region indicated by the horizontal dotted lines).}
    \label{fig:tsz_systematics_test}
\end{figure}

Figure~\ref{fig:tsz_systematics_test} shows a comparison of the mass likelihoods obtained from either the baseline minimum variance map or the more noisy tSZ-nulled minimum variance map, with the two appearing in good agreement. Nulling the tSZ signal significantly increases the map variance, primarily due to the CIB and instrumental noise terms. Thus using the tSZ-nulled map approximately doubles the mass uncertainty compared to the baseline case where the tSZ signal is removed by the random rotation stacking procedure. We choose to proceed with the baseline map. 

The success of the random rotation technique in mitigating tSZ contamination has also been verified using a larger sample of simulated clusters, in previous studies \cite{{Levy2023}, {Raghunathan2019Pol}}. Similarly, we use the simulations described in Section~\ref{sec: Simulations}, to estimate the contribution from residual tSZ and kSZ, as well as foregrounds due to CIB and radio galaxies, to our systematic error. This is done by running our pipeline based on a simulation set without foregrounds, tSZ, and kSZ signals, and then repeating the run on the same simulation set with the tSZ, kSZ, and foregrounds added in. Based on this test, we find a contribution of $1.3\%$ to our error budget, equivalent to a shift of $0.12\sigma$ in our results.

\subsubsection{Cluster miscentering}
\label{sec: miscent_sys}

To estimate the systematic uncertainty due to our cluster miscentering model, we repeat the cluster mass measurement for the Volume\&Redshift-Limited sample with our fiducial setup but change the miscentering parameters by their $1\sigma$ uncertainty to $f_{\textup{mis}}=0.30$ (see \cite{McClintock2019}) and $\ln{c_{\textup{mis}}}=-1.35$. In this case, the measured mean cluster mass increases by $1.7\%$ or $0.15\sigma$.\footnote{We note that if one uses a larger uncertainty on $f_{\textup{mis}}$ (e.g. based on the \cite{Rykoff2016} estimate from the smaller DES Science Verification sample) and takes $f_{\textup{mis}}=0.33$, the result changes by $\sim 3.4\%$ ($\sim 0.29\sigma$).}

\subsubsection{Filtering model}
\label{sec: TF_sys}

As described in Section~\ref{sec: Simulations}, we estimate the impact of filtering applied to our CMB maps using a 2D transfer function with high- and low-pass filter components set to $\ell_1=300$ and $\ell_2=13000$ respectively. While $\ell_2$ is set to angular scales that do not matter to our lensing reconstruction, we need to calculate the systematics due to the uncertainty on the position of the high-pass filter $\ell_1 = 300\pm20$. To this end, we recompute our models assuming $\ell_1=280$ and 320 and evaluate the changes in the mean lensing mass of our Volume\&Redshift-Limited sample. Based on this analysis, we find the systematic contribution of our filtering model to be $0.9\%$ (or $0.07\sigma$).

\subsubsection{Beam uncertainties}
\label{sec: beam_sys}

To estimate the uncertainties in our modelling of the telescope beam profile, we generate 10 Gaussian realisations of the beam and regenerate our models of the lensing dipole (described in Section~\ref{sec: Lensing dipole models}) for these beams. To generate the simulated beams, we take the beam covariance matrix $\Sigma_{\ell\ell'} = \langle \delta B_\ell \delta B_{\ell'} \rangle$, with $\delta B_\ell$ denoting the deviation of $B_\ell$ from its mean $\bar{B}_\ell$. Here, the elements of the covariance matrix are given by $\Sigma_{\ell\ell'} = (Q\Lambda Q^\top)_{\ell\ell'}$, where, the diagonal matrix $\Lambda$ contains the eigenvalues $\lambda_i$ of the covariance matrix, and the orthogonal matrix $Q$, contains the eigenvectors $v_i$ of the covariance matrix, satisfying $\Sigma v_i = \lambda_i v_i$.

Under the assumption that $\Sigma_{\ell\ell'}$ encapsulates the statistical characteristics of the beam (i.e., the relevant coefficients follow a Gaussian distribution), we can simulate beam profiles as follows:

\[
\hat{B}_\ell = \bar{B}_\ell + a_{i\ell} e_{i\ell},
\]where $e_{i\ell}$ represents the $\ell$th component of the $i$th column in $Q\sqrt{\Lambda}$, and $a_i$ are standard Gaussian variables with zero mean and unit variance $\langle a_i a_j \rangle = \delta_{ij}$. 

\begin{figure*}
    \centering
    \includegraphics[width=0.8\textwidth]{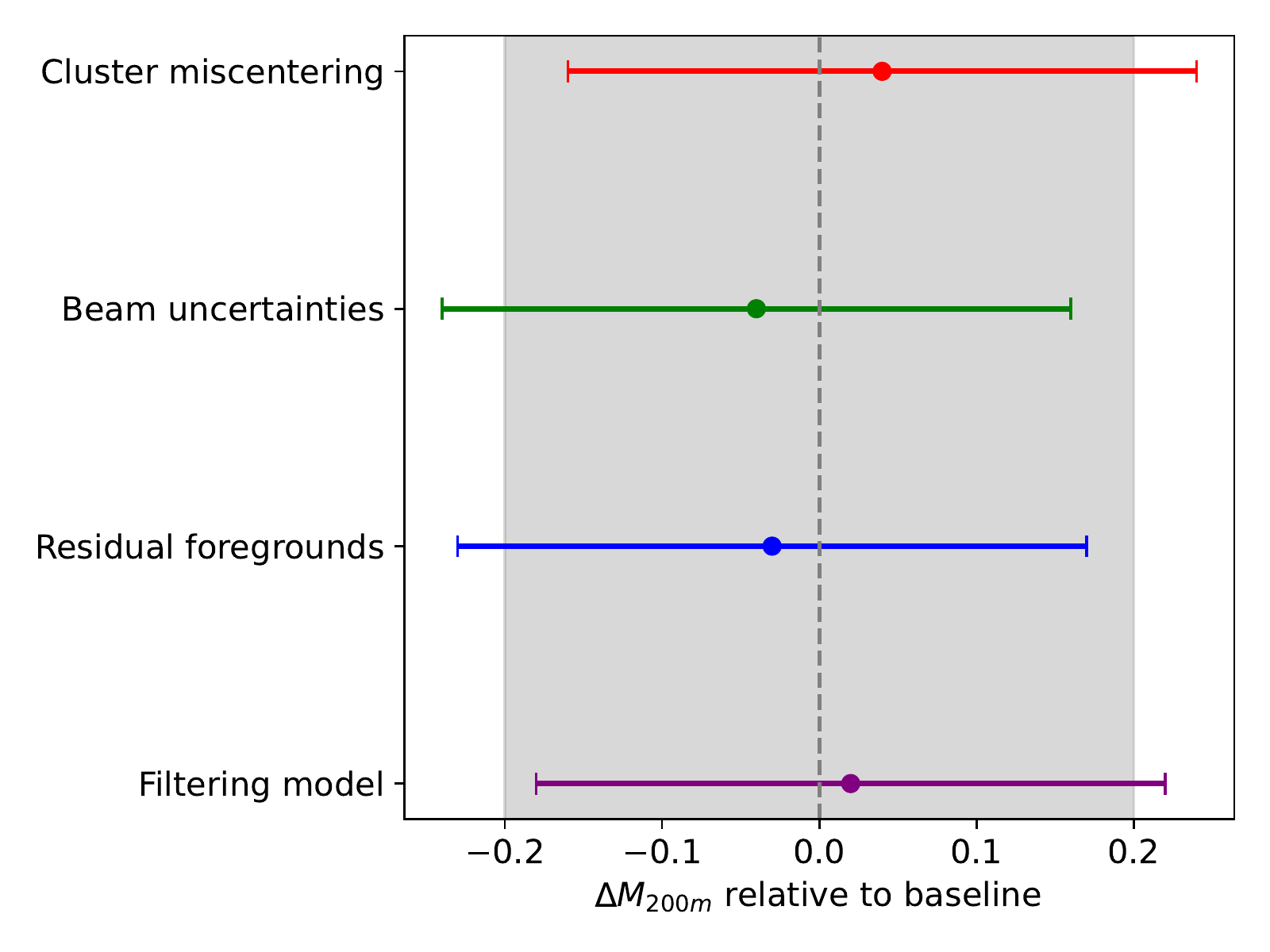}
    \caption{Comparison of the results of systematics tests presented in Table~\ref{tab:systematic_error_budget} relative to the baseline measurement (Volume\&Redshift-Limited sample). The shaded region and error bars indicate the statistical uncertainty of the measurements.}
    \label{fig:sys_tests_plot}
\end{figure*}

We then repeat our measurement of the mean cluster mass for the Volume\&Redshift-Limited sample and take the $1\sigma$ uncertainty on these 10 new measurements as our estimate of the systematic uncertainty due to our beam modelling. Here, we find a $1.7\%$ contribution to our systematic error budget, equivalent to $0.15\sigma$ of our statistical error. We provide a summary of the different contributors to our systematic error budget in Table~\ref{tab:systematic_error_budget} and plot these relative to the baseline measurement in Figure~\ref{fig:sys_tests_plot}.

\begin{table}
\centering
\begin{tabular}{@{}lcc@{}}
\toprule
Source of Error & Magnitude of Error & frac. of $\sigma_{\text{stat}}$ \\ 
\midrule
Cluster miscentering & $1.7\%$ & $0.15\sigma$ \\
Beam uncertainties & $1.7\%$ & $0.15\sigma$ \\
Residual foregrounds & $1.3\%$ & $0.12\sigma$ \\
Filtering model & $0.9\%$ & $0.07\sigma$ \\
Total & $2.4\%$ & $0.25 \sigma$ \\
\bottomrule
\end{tabular}
\medskip \\

\caption{Systematic Error Budget in the Stacked Mass for DES RM Year 3 Volume\&redshift-limited Sample}
\label{tab:systematic_error_budget}
\end{table}

\subsubsection{Optical weak lensing systematics}
\label{sec: optical_WL_sys}

In this work, we perform direct comparisons of our results with cluster masses given by the optical weak lensing calibrated mass--richness scaling relation of \cite{McClintock2019}. As such, we present a brief discussion of systematics impacting optical weak lensing analyses, as well as outstanding cosmological tensions between analyses of the DES Y1 cluster sample and various other cosmological probes.  

The constraints on $\Omega_{\textup{m}}$ and $\sigma_8$ from a joint cluster abundances and weak lensing analysis of the DES Y1 cluster sample are in $2.4\sigma$ tension with the DES Y1 $3\times2$pt results, and in $5.6\sigma$ with the Planck CMB analysis \cite{Abbott2020}. The weak lensing measurements used in this analysis were based on the results of \cite{McClintock2019}, where the systematic uncertainty was estimated to be $4.3\%$. As such, weak lensing mass calibration systematics alone are not sufficient to explain the tensions found in the analysis of \cite{Abbott2020}.

In a later work, \cite{Grandis2021} explored the contamination of the DES-Y1 cluster sample with SPT-SZ selected clusters. Here, it was shown that $10-20\%$ of the $\lambda<40$ DES redMaPPer clusters are galaxy groups with masses of $\sim3-5\times{10}^{13}\,{\rm{M}}_{\odot }$ that are misclassified as more massive clusters due to projection effects. The presence of such low-mass systems in the DES Y1 sample is likely a dominant contributing factor to the cosmological tensions presented in \cite{Abbott2020}. Indeed, it was shown in \cite{Abbott2020}, that tensions can be significantly alleviated by limiting the sample to clusters with $\lambda\geq30$, which further supports this hypothesis. 

It is important to note that systematics due to the presence of such low-mass contaminants will also bias the CMB lensing measurements presented in this study, and in the future, more work is needed to better characterise and minimise the impact of such contaminants via spectroscopic follow-up, use of different cluster detection algorithms, and comparison to future X-ray and SZ samples which probe lower cluster masses.

\subsection{Cluster mass--richness scaling relation}
\label{sec: Mass_richness_scaling_relation}

\begin{figure*}
    \centering
    \includegraphics[width=\textwidth]{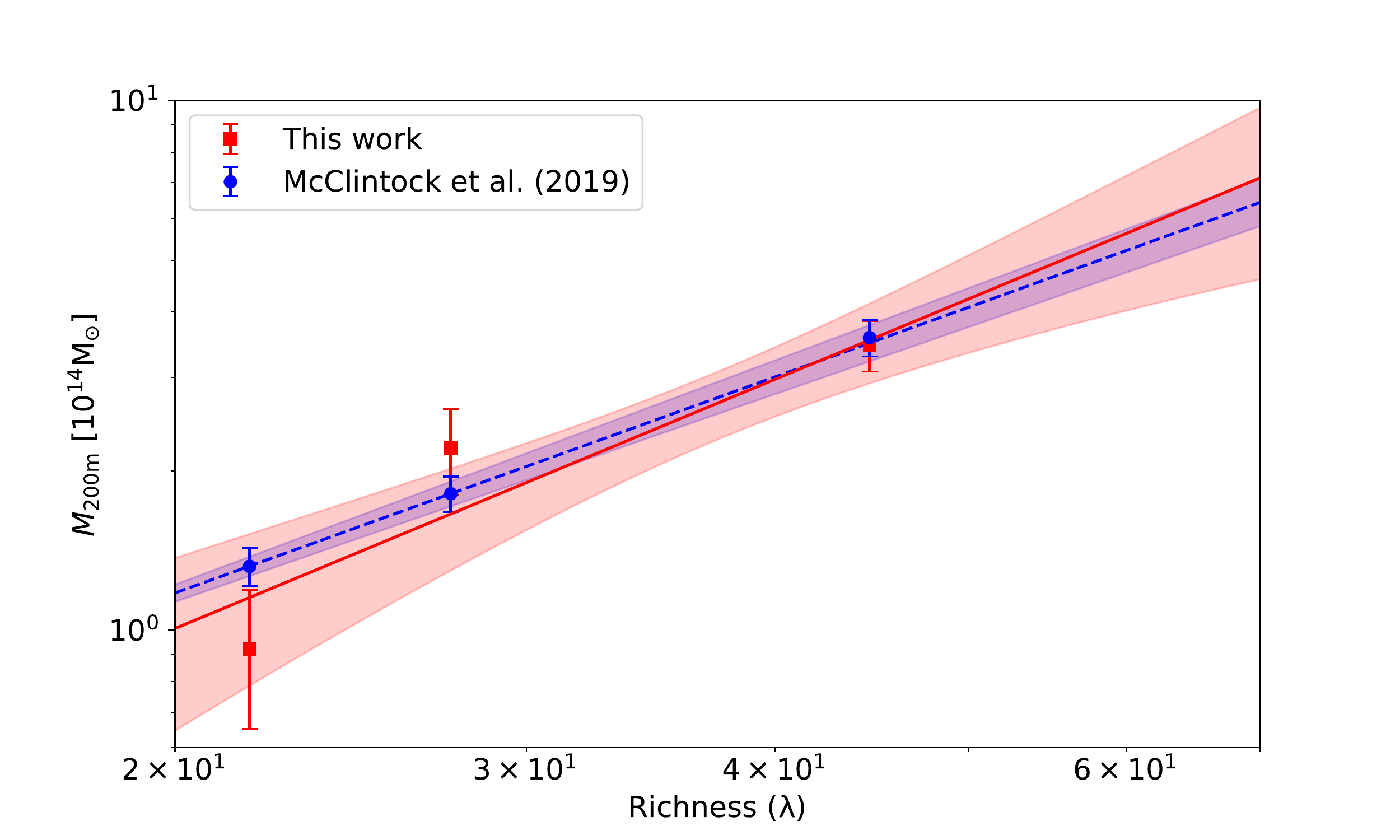}
    \caption{Mass--richness scaling relation fitted to our measured cluster masses of the Volume\&Redshift-Limited sample in three richness bins (red squares/solid line), compared to the cluster masses given by the mass--richness scaling relation of \cite{McClintock2019} (blue circles/dashed line). The shaded regions indicate the $1 \sigma$ uncertainty of the scaling relations.}
    \label{fig:mass_richness_scaling_relation}
\end{figure*}

In Figure~\ref{fig:mass_richness_scaling_relation}, we compare our binned cluster mass measurements with those given by the mass--richness scaling relation of \cite{McClintock2019}. We find good agreement between the CMB cluster lensing and optical weak lensing measurements across all richness bins. Based on fitting a two-parameter model (with no redshift evolution), to the cluster mass measurements in our three richness bins, we obtain a mass--richness scaling relation given by:

\begin{equation}
M_{200m} = [3.0\pm0.4]\times10^{14}\ {\rm M}_\odot (\lambda/40)^{1.6\pm0.5}. 
\label{eq:mass_richness_scaling_relation}
\end{equation}

This best-fit model has a chi-square value of 2.59 ($p$-value of 0.11). We note that the constraining power of the current data limits us to 3 richness bins, which matches the number of free parameters in the \cite{McClintock2019} model. As such, we are unable to perform statistical tests to quantitatively assess the level of agreement between our data and the \cite{McClintock2019} scaling relation which contains 3 free parameters, and leave this to future analyses.

\subsection{Mean cluster mass as a function of redshift}
\label{sec: Mass_richness_relation_bins}

\begin{figure}
    \centering
    \includegraphics[width=0.8\columnwidth]{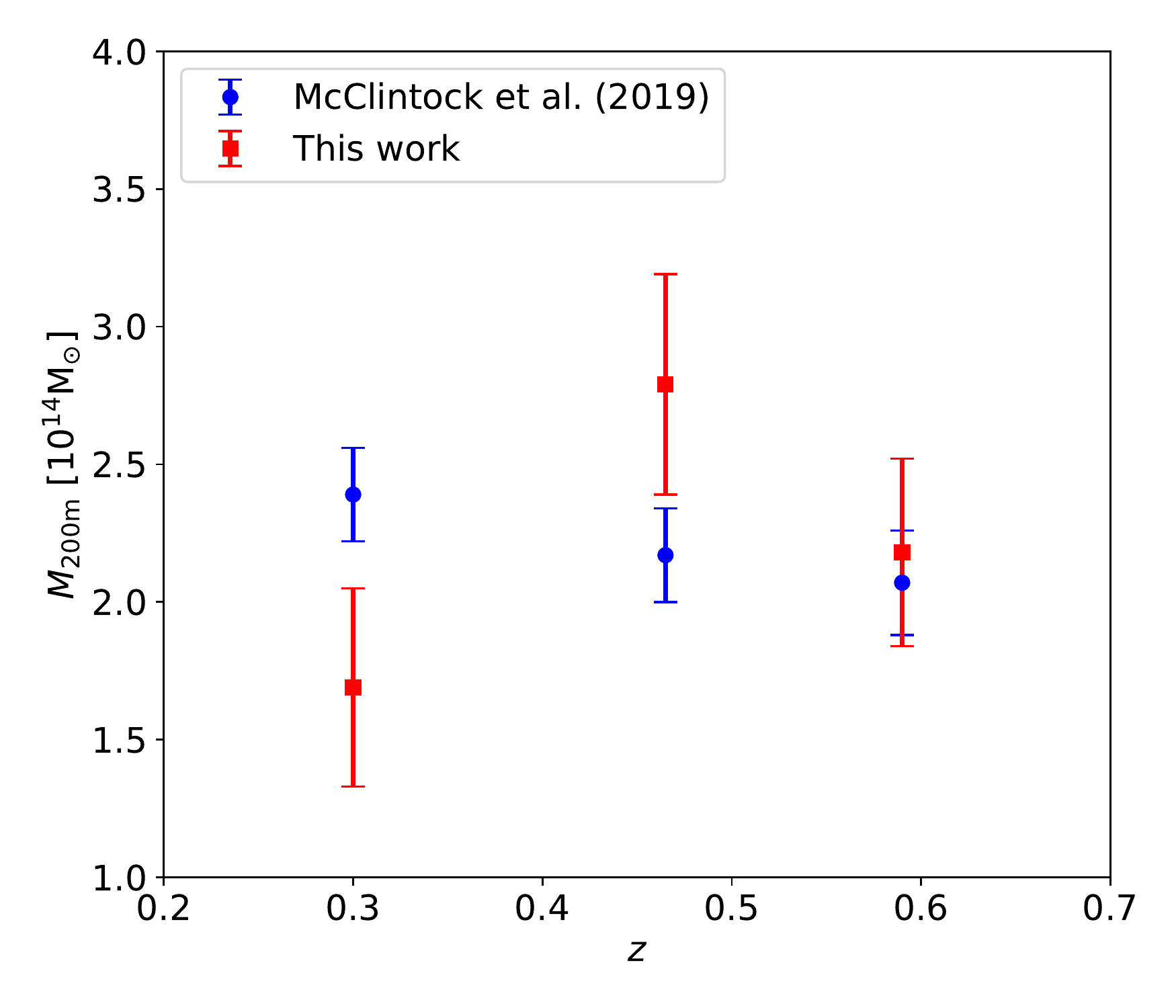}
    \caption{Comparison of the mean cluster mass measurements of this work for three redshift bins, to masses based on the optical weak lensing calibrated, mass--richness scaling relation of \cite{McClintock2019}.}
    \label{fig:vol_lim_likelihoods_bins}
\end{figure}

Figure \ref{fig:vol_lim_likelihoods_bins} shows the mean cluster mass for three redshift sub-samples of our Volume\&Redshift-Limited sample described in Table~\ref{tab:z_lambda_subsamples}. The red data points show the measurements from this work compared to the mean masses of the same subsamples (blue data points), obtained using the optical weak lensing calibrated mass--richness scaling relation of \cite{McClintock2019}. We find a reasonable agreement between the measurements, noting only a $\sim1.7\sigma$ deviation between the two measurements in the lowest redshift bin. As this modest divergence is not statistically significant, we do not investigate it further at this stage. However, it would be interesting to see if such a discrepancy persists in future SPT-3G measurements at a higher level of statistical significance.

\subsection{Forecasts}
\label{sec: Forecasts}

In this section, we provide forecasts for the expected mass constraints one could achieve by including polarization data and upon the completion of the SPT-3G survey, including data from the SPT-3G `Extended' survey (providing a total of $\sim2800$ deg$^2$ overlap with DES, albeit at varying sensitivities). We summarise these forecasts in Table~\ref{tab:Forecasts}. Here, we can see that upon completion of the survey in 2026 and by including polarization and the Extended survey observations, we can improve the current mass constraints by a factor of $\sim1.8$, obtaining a $5.7\%$ stacked cluster mass constraint. This is much more competitive with optical weak lensing mass constraints which currently provide a $\sim5\%$ cluster mass constraint for the DES sample \cite{McClintock2019} and will allow for more precise cosmological parameter estimation. For future, high S/N measurements, under the simplifying assumption that the uncertainties of the CMB cluster lensing and optical weak lensing measurements are uncorrelated, one could expect a $\sim\sqrt{2}$ improvement in the cluster mass constraint upon combining the two measurements. However, combining the two measurements would require detailed analysis of the level of correlation between the joint systematics of the two techniques.

\begin{table}
    \centering
    \begin{tabular}{cccccc}
       \toprule 
        SPT-3G survey field & Data &Years observed& Map depth  & DES overlap  & Mass constraint \\
         &  & & [$\mu$K.arcmin] &  [deg$^2$] &   [$\%$]\\
        \hline
         Main (This work) & T & 2019+20 & 3.1 & $1350$ & $9.8$ \\
         Main & T & 2019-23, 2025-26 & 1.6 & $1350$ & $7.1$ \\
         Main & T+Pol & 2019-23, 2025-26 & 1.6 & $1350$ & $6.4$ \\
         Extended & T+Pol & 2019-23 & 6.1 & $1420$ & $11.8$ \\
         Main+Extended & T+Pol & -- & -- & $2770$ & $5.7$ \\
    \bottomrule
    \end{tabular}
    \caption{Mass constraint forecasts with the addition of various SPT-3G survey data. Here, the map depth is given by the inverse quadrature sum of the noise in the 95, 150 and 220 GHz frequency bands.}
    \label{tab:Forecasts}
\end{table}

\section{Conclusions}
\label{sec: Conclusions}

In this study, we presented a measurement of the mean cluster masses of three DES-Y3 galaxy cluster samples using the CMB cluster lensing measurements from the initial two years of observations of the SPT-3G survey. Here, we restrict our measurements to the temperature data and the `Main' SPT-3G survey which has a $1350$ deg$^2$ overlap with DES (after masking) and leave the addition of SPT-3G polarization maps, as well as data from the SPT-3G `Extended' survey to future works when additional SPT-3G observations are available.

\begin{linenomath*}
The DES-Y3 cluster samples used in this analysis consist of a Flux-Limited sample with 8865 clusters, a Volume-Limited sample with 5391 clusters and a Volume\&Redshift-Limited sample with 4450 clusters. The latter sample is designed to match the selection function of the cluster sample used for DES cluster cosmology analyses, and thus is the primary focus of this work. For the three samples, we detect the CMB lensing dipole with a significance of $12.4\sigma$, $10.5\sigma$ and $10.2\sigma$ and find the mean cluster masses to be:
\end{linenomath*}
\begin{linenomath*}
\begin{align}
    {M}_{200{\rm{m}}} &= 1.66 \pm 0.13 \text{ [stat.]}\pm 0.03 \text{ [sys.]} \times 10^{14}\,\textup{M}_\odot \quad (8.0\%), \nonumber \\
    {M}_{200{\rm{m}}} &= 1.97 \pm 0.18 \text{ [stat.]}\pm 0.05 \text{ [sys.]} \times 10^{14}\,\textup{M}_\odot \quad (9.5\%), \nonumber \\
    {M}_{200{\rm{m}}} &= 2.11 \pm 0.20 \text{ [stat.]} \pm 0.05 \text{ [sys.]} \times 10^{14}\,{\rm{M}}_{\odot } \quad (9.8\%) \nonumber.
\end{align}
\end{linenomath*}

This measurement represents a factor of $\sim2$ improvement in precision relative to CMB cluster lensing measurements based on previous generations of SPT surveys (see, e.g. \cite{Baxter2018}, \cite{Raghunathan2019T} and \cite{Raghunathan2019Pol}) and is much more competitive with optical weak lensing mass constraints. Overall, we find good agreement between our measurements and those given by the redMaPPer mass--richness scaling relations of previous works (e.g. \cite{Farahi2016}, \cite{Geach2017}, \cite{Melchior2017}, \cite{Baxter2018}, \cite{Raghunathan2019T} and \cite{McClintock2019}) calibrated using techniques including CMB cluster lensing, optical weak-lensing, and velocity dispersion measurements from various combinations of DES, SDSS, and Planck data. 

We verify that our measurements are not significantly biased due to contamination from the residual tSZ signal by comparing the mass of our Volume\&Redshift-Limited sample to the mass of the same sample obtained using CMB cluster lensing of a tSZ-nulled ILC map. We find a good agreement between the two measurements, with the higher noise levels in the tSZ-nulled map resulting in a factor of $\sim2$ larger uncertainty on the mean cluster mass. 

Although it was not possible to measure cluster masses for different redshift and richness bins in previous SPT works due to the low S/N of the lensing dipole, the improved sensitivity of the SPT-3G data and the greater overlap with DES enables us to divide our Volume\&Redshift-Limited sample into 3 redshift and richness bins, each containing $\sim1/3$ of the clusters in the full sample. For these subsamples, we obtain mass constraints ranging from $10-20\%$ in precision (with a mean precision of $14\%$). Our results across these sub-samples do not reveal any significant discrepancies when compared to the optical weak lensing calibrated masses given by the scaling relation of \cite{McClintock2019}, although we observe that our mass measurement in the lowest redshift bin is $\sim1.7\sigma$ lower.

Finally, we perform forecasts for expected mass constraints using various combinations of the upcoming SPT-3G data. We find that upon completion of the survey in 2026, the combination of temperature and polarization data would yield mass constraints of $6.4\%$ and $11.8\%$ in the Main and Extended SPT-3G surveys, respectively, which translates to a $5.7\%$ mass constraint upon the combination of the full SPT-3G datasets. This level of precision will make CMB cluster lensing much more competitive with optical weak lensing in future years. Moreover, it will serve as an important test of the robustness of cluster mass measurements for precision cosmology due to the complementary systematics of the two measurements.


\acknowledgments

We thank the anonymous referee for their review of the paper and useful comments. 

The Melbourne team acknowledges support from the Australian Research Council’s Discovery Projects scheme (DP200101068). 

SR acknowledges support from the Center for AstroPhysical Surveys (CAPS) at the National Center for Supercomputing Applications (NCSA), University of Illinois Urbana-Champaign. 

The South Pole Telescope program is supported by the National Science Foundation (NSF) through award OPP-1852617. Partial support is also provided by the Kavli Institute of Cosmological Physics at the University of Chicago.

This project has received funding from the European Research Council
(ERC) under the European Union’s Horizon 2020 research and innovation
programme (grant agreement No 101001897)

Funding for the DES Projects has been provided by the U.S. Department of Energy, the U.S. National Science Foundation, the Ministry of Science and Education of Spain, the Science and Technology Facilities Council of the United Kingdom, the Higher Education Funding Council for England, the National Center for Supercomputing Applications at the University of Illinois at Urbana-Champaign, the Kavli Institute of Cosmological Physics at the University of Chicago, the Center for Cosmology and Astro-Particle Physics at the Ohio State University, the Mitchell Institute for Fundamental Physics and Astronomy at Texas A\&M University, Financiadora de Estudos e Projetos, Funda{\c c}{\~a}o Carlos Chagas Filho de Amparo {\`a} Pesquisa do Estado do Rio de Janeiro, Conselho Nacional de Desenvolvimento Cient{\'i}fico e Tecnol{\'o}gico and the Minist{\'e}rio da Ci{\^e}ncia, Tecnologia e Inova{\c c}{\~a}o, the Deutsche Forschungsgemeinschaft and the Collaborating Institutions in the Dark Energy Survey. 

The Collaborating Institutions are Argonne National Laboratory, the University of California at Santa Cruz, the University of Cambridge, Centro de Investigaciones Energ{\'e}ticas, Medioambientales y Tecnol{\'o}gicas-Madrid, the University of Chicago, University College London, the DES-Brazil Consortium, the University of Edinburgh, the Eidgen{\"o}ssische Technische Hochschule (ETH) Z{\"u}rich, Fermi National Accelerator Laboratory, the University of Illinois at Urbana-Champaign, the Institut de Ci{\`e}ncies de l'Espai (IEEC/CSIC), the Institut de F{\'i}sica d'Altes Energies, Lawrence Berkeley National Laboratory, the Ludwig-Maximilians Universit{\"a}t M{\"u}nchen and the associated Excellence Cluster Universe, the University of Michigan, NSF's NOIRLab, the University of Nottingham, The Ohio State University, the University of Pennsylvania, the University of Portsmouth, SLAC National Accelerator Laboratory, Stanford University, the University of Sussex, Texas A\&M University, and the OzDES Membership Consortium.

Based in part on observations at Cerro Tololo Inter-American Observatory at NSF's NOIRLab (NOIRLab Prop. ID 2012B-0001; PI: J. Frieman), which is managed by the Association of Universities for Research in Astronomy (AURA) under a cooperative agreement with the National Science Foundation.

The DES data management system is supported by the National Science Foundation under Grant Numbers AST-1138766 and AST-1536171. The DES participants from Spanish institutions are partially supported by MICINN under grants ESP2017-89838, PGC2018-094773, PGC2018-102021, SEV-2016-0588, SEV-2016-0597, and MDM-2015-0509, some of which include ERDF funds from the European Union. IFAE is partially funded by the CERCA program of the Generalitat de Catalunya. Research leading to these results has received funding from the European Research Council under the European Union's Seventh Framework Program (FP7/2007-2013) including ERC grant agreements 240672, 291329, and 306478. We acknowledge support from the Brazilian Instituto Nacional de Ci\^encia e Tecnologia (INCT) do e-Universo (CNPq grant 465376/2014-2).

This manuscript has been authored by Fermi Research Alliance, LLC under Contract No. DE-AC02-07CH11359 with the U.S. Department of Energy, Office of Science, Office of High Energy Physics.

Argonne National Laboratory’s work was supported by the U.S. Department of Energy, Office of High Energy Physics, under Contract No. DE- AC02-06CH11357.

We acknowledge the use of \url{http://astromap.icrar.org/} for producing Fig.~\ref{fig:Survey_footprints}. Some of the results in this paper have been derived using the \textsc{healpy} \cite{healpy} and \textsc{HEALPix} \cite{healpix} packages. The analysis of this work also made use of \textsc{Python3} \cite{python3} and packages including \textsc{scipy} \cite{Scipy}, \textsc{numpy} \cite{NumPy}, \textsc{astropy} \cite{astropy2018} and \textsc{pandas} \cite{pandas}.


\bibliographystyle{JHEP}
\bibliography{bibliography}



\end{document}

%% file: spt3g_des_cmb_cluster_lensing_authours.tex
\author[\Melbourne]{B.~Ansarinejad,}
\author[\ILNCSA]{S.~Raghunathan,}
\author[\CTIO]{T.~M.~C.~Abbott,}
\author[\Cardiff]{P.~A.~R.~Ade,}
\author[\LIneA]{M.~Aguena,}
\author[\UMICH]{O.~Alves,}
\author[\KICPChicago,\AAUChicago,\FNAL]{A.~J.~Anderson,}
\author[\UMICH]{F.~Andrade-Oliveira,}
\author[\ILAst,\ILNCSA]{M.~Archipley,}
\author[\IAP]{L.~Balkenhol,}
% ...
